\documentclass[aps,prl,twocolumn,superscriptaddress,oneside,floatfix,amsmath,showpacs,amssymb]{revtex4}
\usepackage{graphicx}
\usepackage{dcolumn}
\usepackage{bm}

\begin{document}

\newcommand{\Y}{YBa$_2$Cu$_3$O$_y$}
\newcommand{\ie}{\textit{i.e.}}
\newcommand{\eg}{\textit{e.g.}}
\newcommand{\etal}{\textit{et al.}}


\title{Reply to Comment on ``Onset of boson mode at the superconducting critical point of underdoped \Y''}


\author{Nicolas~Doiron-Leyraud}
\affiliation{D\'epartement de physique and RQMP, Universit\'e de Sherbrooke, Sherbrooke, Canada}

\author{Louis~Taillefer}
\email{Louis.Taillefer@USherbrooke.ca}
\affiliation{D\'epartement de physique and RQMP, Universit\'e de Sherbrooke, Sherbrooke, Canada}
\affiliation{Canadian Institute for Advanced Research, Toronto, Canada}

\date{\today}

\pacs{74.25.Fy, 74.72.Bk, 72.15.Eb}
\maketitle


In our recent Letter \cite{Doiron-Leyraud06}, we reported a study of how the thermal conductivity of \Y (YBCO) changes as its doping is made to cross the superconducting critical point at $p_c=0.05$, \ie, from $p_i<p_c$ to $p_f>p_c$. For the same crystal with the same contacts, we showed that the difference $\Delta\kappa=\kappa(p_i)-\kappa(p_f)=\beta T^3$ up to 500 mK. This is an experimental fact, free from any analysis and confirmed on several specimens. We attribute this difference to heat-carrying bosons with a $T^3$ conductivity analogous to the magnons in the undoped cuprate antiferromagnet Nd$_2$CuO$_4$ \cite{LiMagnon}.


\begin{figure}[h!]
\centering
\includegraphics[scale=0.47]{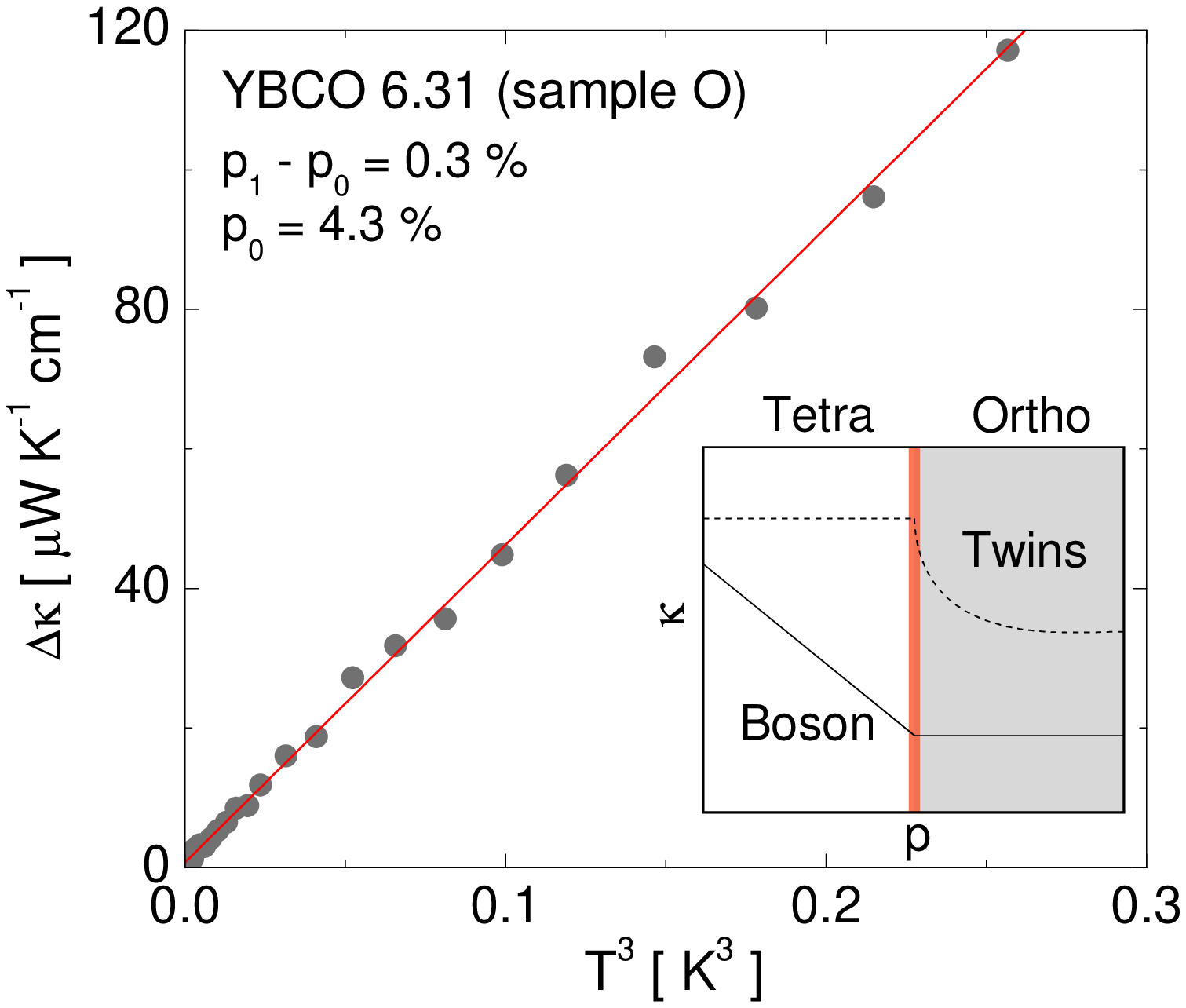}
\includegraphics[scale=0.47]{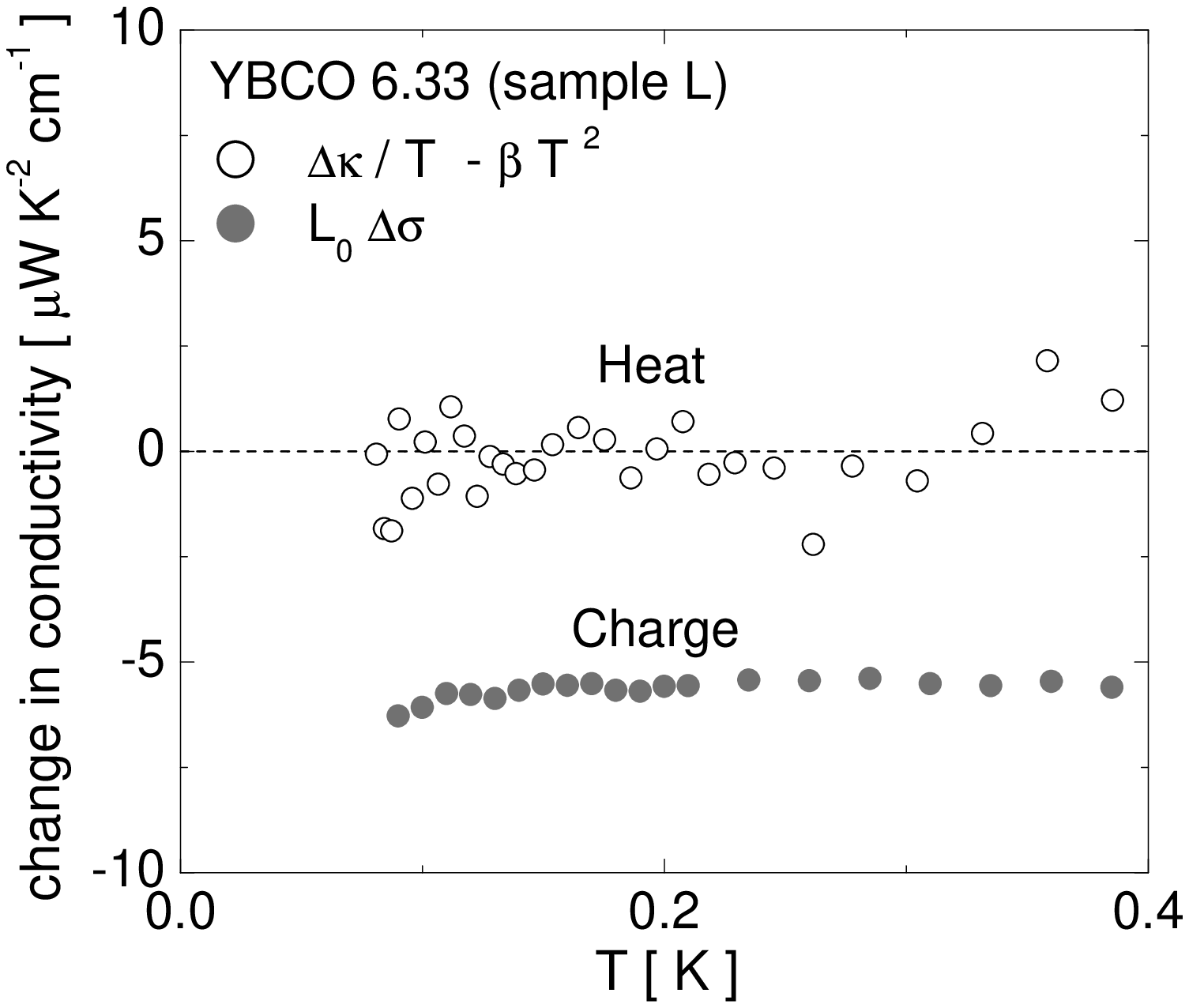}
\caption{\textbf{Top:} Boson - Change in the thermal conductivity $\kappa$ of YBCO sample O (y = 6.31) between $p_0 = 4.3$\% and $p_1 = 4.6$\%, well below the I-S transition at $p = 5$\%. The reduction in $\kappa$, giving $\Delta\kappa \propto T^3$, occurs entirely within the tetragonal phase and cannot be due to scattering off twin boundaries. Inset: sketch of how $\kappa$ would evolve if twin boundary scattering of phonons were causing a reduction (dashed line) and how $\kappa$ is observed to change in our measurements (solid line)(see Fig. 2 in \cite{Doiron-Leyraud06}). \textbf{Bottom:} - Change in $\kappa$ of YBCO sample L (y = 6.33) between $p_0 = 4.7$\% and $p_1 = 5.0$\%, with the $\beta T^3$ contribution subtracted, plotted as $\Delta\kappa/T - \beta T^2$ vs $T$. The corresponding change in electrical conductivity $\sigma$ (see Fig. 3 of \cite{Doiron-Leyraud06}) is plotted as $L_0\Delta\sigma$ vs $T$.}
\label{figAB}
\end{figure}


\textit{Twin-boundary scattering.} -- Ando \cite{Andocondmat} points out correctly that YBCO undergoes a structural transition from a tetragonal (tetra) phase for $p<p_c$ to an orthorhombic (ortho) phase for $p>p_c$. He raises the possibility that in twinned samples like ours, scattering from twin boundaries will cause a drop in phonon conductivity when crossing into the ortho phase. While this may be true, it does not explain the $\beta T^3$ difference we observe because phonon conductivity in (twinned or untwinned) cuprate crystals is never $T^3$, at any doping. This was shown very clearly in our study of Nd$_2$CuO$_4$ \cite{LiMagnon}, where the magnon conductivity goes precisely as $T^3$, while that of phonons goes as $T^{2.6}$. Below 0.5 K or so, the mean free path of magnons is limited by the rough sample edges (quasi-2D) while that of phonons is limited by the sample faces (3D) which, in as-grown crystals, are mirror-like and cause specular reflection. This introduces a strong $T$-dependence to the phonon mean free path, which means that $\kappa_p$ is no longer cubic in $T$, as shown in our recent study of as-grown and roughened samples of Nd$_2$CuO$_4$ \cite{LiRough}. Therefore, the $T^3$ term we observe in YBCO cannot be due to a change in $\kappa_p$ caused by additional phonon scattering for $p>p_c$. If significant at all, this additional scattering will be seen at higher temperature, above the boundary scattering regime ($T>0.5$ K or so). (Note that in their own measurements on YBCO, Ando and co-workers claim to see a phonon $T^3$ term at low $T$ \cite{Sun}. However, this is always limited to temperatures below 150 mK or so.)

Although their data is not shown in our Letter \cite{Doiron-Leyraud06}, we have measured samples whose entire doping evolution took place in the tetra phase, \ie, below 5\%. We refer to these samples (labeled M, N, O) in our Letter and in Fig.\ref{figAB} we show data for the most underdoped one (sample O, with y = 6.31). The change in conductivity between $p_0=4.3$\% and $p_1=4.6$\% exhibits the same $T^3$ dependence shown in our Letter \cite{Doiron-Leyraud06}. Because sample O never entered the ortho phase, twin boundaries cannot be invoked as the cause of the reduction in $\kappa$. Invoking strains that build up as the tetra--ortho transition is approached from below also does not work, as scattering from strain fields is known to vary linearly with phonon frequency and so causes $\kappa_p \propto T^2$, not $T^3$.

\textit{Wiedemann-Franz law.} -- In our Letter, we fit the measured $\kappa(T)$ to a sum of three terms: $\kappa(T) = aT +  \beta T^3 + r(T)$ coming from fermions, bosons, and phonons, respectively. Ando argues that we should have explicitly included an additional term related to charge conduction. However, the only way one can do this \textit{a priori} is to use the Wiedemann-Franz (WF) law: $\kappa_e = L_0 T \sigma = L_0 T / \rho$, where $L_0$ is the Sommerfeld constant and $\sigma$ the electrical conductivity. In other words, Ando wants us to replace $\kappa_f = aT$ by $\kappa_f = a' T + L_0 T / \rho$.

Moving away from arguments over fitting procedure, the best way to compare charge and heat conductivities is to directly examine the respective changes induced by doping. In the bottom panel of Fig.\ref{figAB}, we plot the difference in $\kappa$ for sample L between $p_0 = 4.7$\% and $p_1 = 5.0$\%, without the $T^3$ term, \ie, $\Delta\kappa/T - \beta T^2$ vs $T$. We compare this with the corresponding change (on the same sample at the very same dopings) in electrical conductivity, $L_0 \Delta\sigma$ \cite{Doiron-Leyraud06}. Because $\kappa/T$ does not change with doping while $\sigma$ does, the WF law is unambiguously violated, independently of the fitting precedure. This shows that Ando's suggestion of including an explicit $\kappa_e$ term leads to an inconsistency, in the sense that it relies on the WF law which does not hold here.


\end{document}